\documentclass[pdflatex,sn-mathphys-num]{sn-jnl}

\usepackage{graphicx}%
\graphicspath{{figs/}}%
\usepackage{multirow}%
\usepackage{amsmath,amssymb,amsfonts}%
\usepackage{amsthm}%
\usepackage{mathrsfs}%
\usepackage[title]{appendix}%
\usepackage{xcolor}%
\usepackage{textcomp}%
\usepackage{manyfoot}%
\usepackage{booktabs}%
\usepackage{url}%

\newcommand{\TD}{T_{\Delta}} 

\begin{document}

\title[Single-visit screening for intracochlear tissue growth]{Single-visit screening for intracochlear tissue growth validated against multi-visit electrical measurement changes}
\author*[1]{\fnm{Minh} \sur{Vu}}\email{minh.q.vu@vanderbilt.edu}
\author[2]{\fnm{Erin} \sur{Bratu}}\email{erin.bratu@univ-rennes1.fr}
\author[1]{\fnm{Jared} \sur{Rybarczyk}}\email{jared.a.rybarczyk@vanderbilt.edu}
\author[1]{\fnm{Jack} \sur{Noble}}\email{jack.noble@vanderbilt.edu}

\affil*[1]{%
\orgdiv{Department of Electrical and Computer Engineering},
\orgname{Vanderbilt University},
\orgaddress{\city{Nashville}, \state{TN}, \country{USA}}}

\affil[2]{%
\orgdiv{INSERM, LTSI -- UMR 1099},
\orgname{University of Rennes},
\orgaddress{\postcode{35000}, \city{Rennes}, \country{France}}}

\abstract{\textbf{Purpose:} Tissue growth (TG) around cochlear implant electrodes raises impedance but is not reliably visible on CT scans. Routinely recorded stimulus current-induced non-stimulating electrode voltages (SCINSEVs) carry this information, but a single recording has no patient-specific reference. We present an algorithm for detecting signs of significant TG from a single SCINSEV measurement.

\textbf{Methods:} TG is assumed to concentrate basally near the electrode insertion site, leaving the apical region mostly unaffected. A patient-specific forward model built from CT is fitted to the apical measurements and predicts the basal ones. The misfit score $T$ measures each basal contact's signed relative misfit against the apical misfit, accounting for apical fit quality and measurement noise. An ear is flagged when two adjacent basal contacts cross the threshold, set for $\leq$5\% false positives on intra-operative recordings assumed growth-free (97 ears). We compare $T$ with $\TD$, a model-free reference tracking basal relative to apical SCINSEV change since the intra-operative recording, persisting over the last two visits (68 ears).

\textbf{Results:} $T$ had a false-positive rate of 3\% on growth-free ears. It agreed with $\TD$ in 81\% of the 68 ears (sensitivity 71\%; specificity 86\%; $p = 6\times10^{-6}$). Over 200 patient-level two-fold splits, median held-out agreement was 78\% at a false-positive rate of 5.1\%.

\textbf{Conclusion:} An apically fitted patient-specific model predicts basal SCINSEVs well enough to flag signs of significant TG at a single visit, without an intra-operative baseline. Detections largely agree with persistent multi-visit basal-versus-apical change, a surrogate ground truth for significant TG.}


\maketitle

\section{Introduction}

Cochlear implants (CIs) restore hearing by electrically stimulating the auditory nerve in individuals with severe to profound hearing loss, but outcomes vary widely between recipients~\cite{ci-outcomes}. One contributor that is rarely observable in a living patient is the tissue response to the implant itself. Following insertion, fibrous tissue and, in some ears, new bone form around the electrode array~\cite{somdas2007,foggia2019}. This tissue growth (TG) is typically most extensive in the basal turn, near the insertion site~\cite{somdas2007,ishai2017}, has been associated with poorer speech recognition~\cite{kamakura2016}, and is accompanied by increased electrode impedance~\cite{wilk2016}.

Detecting TG in a living ear is difficult. Fibrous tissue adjacent to the array is not reliably resolved on clinical CT, where metal artifact from the contacts dominates the region of interest~\cite{ct-tg}. Furthermore, histology is available only post mortem. Routine electrical measurements offer an alternative. When one contact delivers current, the remaining, non-stimulating contacts record voltages that depend on the conductivity of the surrounding tissue. These stimulus current-induced non-stimulating electrode voltages (SCINSEVs)~\cite{derijk2020} rise near a stimulating contact when growth raises the resistivity around it. SCINSEVs can be recorded during routine clinical visits with little added time, and have been used to detect electrode abnormalities such as extracochlear contacts~\cite{derijk2020}.

The difficulty is the absence of a reference. SCINSEVs vary between patients with cochlear anatomy, array position and insertion depth, so an elevated voltage in a single recording cannot be judged without knowing what that particular ear should produce. Comparing against an intra-operative recording supplies a patient-specific reference, but such a baseline is not always available. Confirming that a change persists requires several visits, which delays detection.

Patient-specific electro-anatomical models (EAMs) offer a reference of a different kind. An EAM is a digital twin of the implanted cochlea, built from CT and fitted to the patient's SCINSEVs. It estimates the voltage each contact produces along the auditory nerve for downstream tasks such as neural health estimation~\cite{eam-prior,bratu-lineage,vu-eam,neural-health}. Fitted to the whole recording, however, it would partly absorb the growth it is meant to detect. We therefore use an intentionally lopsided twin. Because TG concentrates basally, it is fitted only to the SCINSEVs evoked from apical contacts, which assumed to be relatively growth-free. We can then sample the simulated potential distribution from basal stimulation sources. TG appears as basal measurements that exceed the prediction by more than the model's error elsewhere allows, which we quantify per basal contact as a misfit score $T$ to flag an ear from a single recording.

However, validation of our method is difficult as no ground truth for TG exists in living patients. We therefore compare $T$ against $\TD$, a model-free reference built from repeated measurements. $\TD$ flags ears whose basal SCINSEVs have risen relative to their apical ones since surgery and remained elevated over the last two visits. Because it draws on the intra-operative baseline and earlier visits, which $T$ does not use, it serves as a surrogate ground truth for significant TG. Across 117 patients, $T$ is calibrated on 97 growth-free recordings and compared with $\TD$ on 68 ears with sufficient follow-up.

\section{Methods}

\subsection{Data}
\label{sec:data}

This retrospective study included 117 patients, each with one implanted ear, a postoperative CT, and SCINSEV recordings acquired at surgery and/or at routine follow-up visits. All arrays had 16 contacts (Advanced Bionics), numbered from contact 1 at the apex to contact 16 at the base. A recording is a $16\times16$ matrix $\mathbf{Z}$, in which $Z_{ij}$ is the voltage recorded at contact $j$ when contact $i$ stimulates, normalized by the delivered current. 
Diagonal entries are excluded throughout.

Because surgery dates were not available separately, the earliest dated recording of each ear was taken to be intra-operative. Contacts flagged during model preparation for localization error, disagreement between CT and recording, or manual review (6 contacts in 5 ears) were excluded from all analyses.

Three sets of recordings were used. The \emph{growth-free arm} comprised the intra-operative recordings of the 97 ears for which a usable recording and model fit were available. The \emph{evaluation set} comprised the 68 ears with an intra-operative recording and at least two follow-up visits, the last at least 30 days after an earlier one. Median follow-up was 557 days (range 105–1099), with a median of 3 follow-up visits. The \emph{same-day null} comprised 145 pairs of recordings made on the same day in 64 ears; each pair consists of two measurements of an ear that cannot have changed between them. 

Each ear's contacts were divided into apical and basal sets from its own CT. Let $\alpha_{\max}$ be the angular insertion depth of the deepest contact. Contacts inserted deeper than $\alpha_{\max}/2$ were classed as apical and the remainder as basal. Insertion depth ranged from 253$^\circ$ to 624$^\circ$ (median 419$^\circ$), and the rule gave 6--9 apical contacts per ear.

\begin{figure}[t]
\centering
\includegraphics[width=0.9\textwidth]{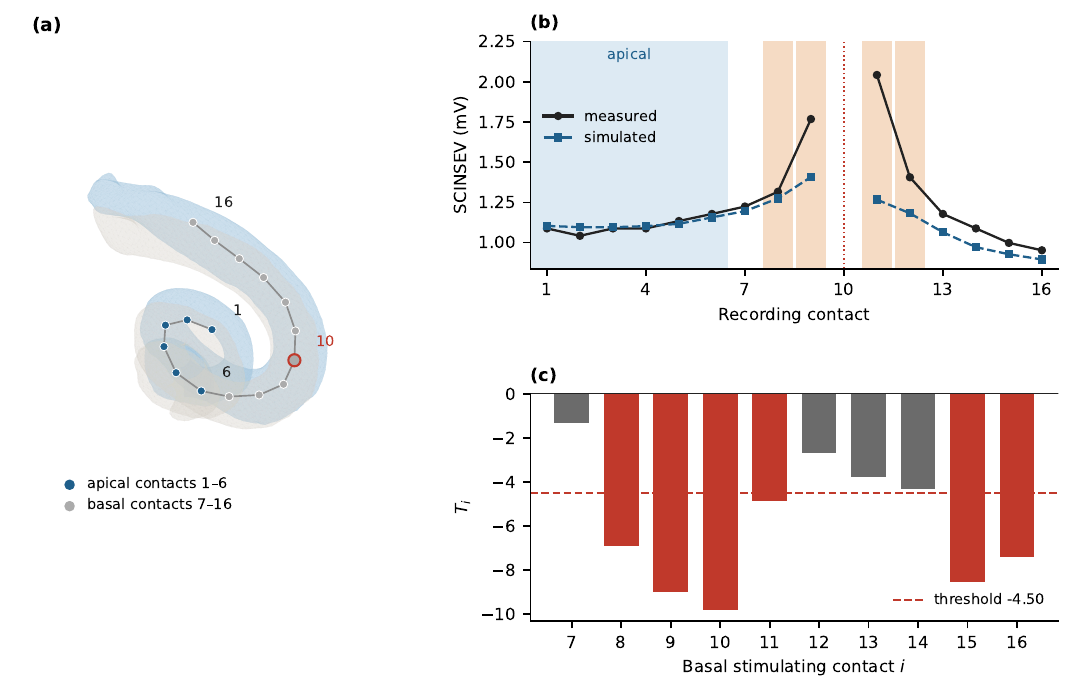}
\caption{Method overview on one ear flagged by both $T$ and $\TD$. (a) Scala tympani, scala vestibuli and electrode array, with apical (blue) and basal (grey) contacts; contact 10 is circled. (b) Measured and simulated SCINSEVs for stimulation at contact 10, with the apical (blue) and neighboring (orange) recording contacts shaded. (c) $T_i$ for each basal contact; contacts below the threshold (dashed) are shown in red.}\label{fig:concept}
\end{figure}

\subsection{Patient-specific EAM and apical fit}
\label{sec:model}

Each ear's EAM was constructed and fitted as described previously~\cite{vu-eam}. Tissue labels for perilymph (scala tympani and vestibuli), soft tissue, modiolus and bone were obtained by atlas-based segmentation of the postoperative CT~\cite{atlas-seg}. The array was localized in the same image~\cite{elec-localization}, and its silicone carrier modeled as insulating. Steady-state current flow was solved by the finite-difference method on a $101\times101\times61$ grid with 0.18~mm voxels. Current injection and readout were distributed trilinearly at the sub-voxel contact positions, producing a simulated matrix $\mathbf{Z}^{\mathrm{mod}}$. Perilymph and soft-tissue resistivities were fitted, bounded to 5--300 and 100--1000~$\Omega\cdot$cm; modiolar resistivity was fixed at 300~$\Omega\cdot$cm and bone linked to soft tissue at 20 times its resistivity~\cite{vu-eam}.

Here the fit used only apical stimulation. The objective was the root-mean-square difference between $\mathbf{Z}^{\mathrm{mod}}$ and the measurement over the rows of the apical contacts, taken across all recording contacts and averaged across those rows. It was minimized by Gaussian-process Bayesian optimization~\cite{bayesopt} over the two resistivities, with 20 initial and 40 sequential evaluations. The fitted model was then solved with stimulation at each basal contact. This gives sampled potential distribution at basal recording sites.

The model was implemented in Python with GPU sparse linear algebra via CuPy~\cite{cupy}. An apical fit took a median of 2.5 minutes (NVIDIA RTX A6000).

\subsection{Single-visit misfit score $T$}
\label{sec:T}

For a basal stimulating contact $i$, let $d_{ij} = Z^{\mathrm{mod}}_{ij} - Z^{\mathrm{meas}}_{ij}$ be the misfit at recording contact $j$, and for a set $\mathcal{J}$ of recording contacts define the relative misfit
\begin{equation}
\delta_i(\mathcal{J}) = \frac{\sum_{j\in \mathcal{J}} d_{ij}}{\sum_{j\in \mathcal{J}} Z^{\mathrm{meas}}_{ij}} .
\label{eq:relmisfit}
\end{equation}
Two sets are used: the near field $N(i)=\{i-2,i-1,i+1,i+2\}$, restricted to valid contacts, and the apical set $A$. The score for contact $i$ is
\begin{equation}
T_i = \frac{\delta_i\bigl(N(i)\bigr) - \delta_i(A)}{\max\bigl(\sigma_i^{A},\, \varepsilon_{\mathrm{rec}}\bigr) \,/\, \bar{Z}_i^{A}} ,
\label{eq:T}
\end{equation}
where $\sigma_i^{A}$ is the standard deviation of $d_{ij}$ over $j\in A$, $\bar{Z}_i^{A}$ the mean of $Z^{\mathrm{meas}}_{ij}$ over $j\in A$, and $\varepsilon_{\mathrm{rec}}$ the reciprocity residual of the recording, with $n$ the number of contacts,
\begin{equation}
\varepsilon_{\mathrm{rec}} = \frac{2}{n(n-1)}\sum_{r<s}\bigl|Z^{\mathrm{meas}}_{rs}-Z^{\mathrm{meas}}_{sr}\bigr| .
\label{eq:recip}
\end{equation}
$T_i$ was computed only when at least three apical recordings in row $i$ were valid.

The numerator is the near-field misfit in excess of the apical misfit of the same row, so error common to the row (such as a global resistivity change or a scale error in the model) largely cancels. The denominator is the spread of the apical misfit, bounded below by the measurement's asymmetry. Because the model is reciprocal, departures from $Z_{rs}=Z_{sr}$ set a scale below which misfit cannot be distinguished from noise~\cite{vu-eam}. Both are relative to the measured magnitude, making $T$ comparable across ears. $T$ is signed: growth raises the measured voltages near contact $i$ above the prediction, making $T_i$ negative.

\begin{equation}
S = \min_{i,\,i+1\,\in\,\mathcal{B}} \max\bigl(T_i,\, T_{i+1}\bigr),
\label{eq:S}
\end{equation}
An ear is flagged when two adjacent basal contacts both score below a threshold $\tau$. Equivalently, with $S$ [Eq. 4] being the lowest value that two adjacent valid basal contacts $\mathcal{B}$ both reach, the ear is flagged if $S<\tau$; $S$ serves as the continuous score in calibration and evaluation. Requiring two adjacent contacts reflects the spatial extent of growth and guards against a single aberrant contact. For each ear, $T$ was computed on one recording from its most recent visit.

The threshold $\tau$ was set on the growth-free arm alone, without reference to $\TD$. Each of the 97 intra-operative recordings was fitted and scored as above, and $\tau$ was the first value on a grid of step 0.25, descending from $-0.25$, at which no more than 5\% of growth-free ears had $S<\tau$.

\subsection{Multi-visit reference $\TD$}
\label{sec:TD}
$\TD$ measures how the basal SCINSEVs of an ear have changed since the intra-operative recording, relative to the change at its apical contacts, without any model. For a follow-up recording $\mathbf{Z}^{(k)}$ and the intra-operative recording $\mathbf{Z}^{(0)}$, the change near contact $i$ is $c_i^{(k)}=\operatorname{mean}_{j\in N(i)}\bigl(Z^{(k)}_{ij}-Z^{(0)}_{ij}\bigr)$, and likewise $c_a^{(k)}$ for each apical contact $a$. Then we have:
\begin{equation}
T_{\Delta,i}^{(k)} = \frac{c_i^{(k)}}{\max\Bigl(\sqrt{\operatorname{mean}_{a\in A}\bigl(c_a^{(k)}\bigr)^2},\; \varepsilon_{\mathrm{rec}}^{(0)},\; \varepsilon_{\mathrm{rec}}^{(k)}\Bigr)} ,
\label{eq:TD}
\end{equation}
being the basal change in units of the typical apical change, or of the reciprocity residual when that is larger. $\TD$ is signed; a rise is the direction of growth.

A contact was positive at a visit if $T_{\Delta,i}^{(k)}\geq\tau_\Delta$. To require persistence, a contact counted only if positive at both the last visit and the most recent earlier visit at least 30 days before it. An ear was positive if two adjacent basal contacts counted. Equivalently, the ear's $\TD$ is the largest, over adjacent basal pairs, of the smallest of the four contact values at those two visits, and the ear is positive if $\TD\geq\tau_\Delta$. The threshold $\tau_\Delta$ was set on the same-day null, scoring each pair as if one recording followed the other. It was the first value on a grid of step 0.25, ascending from 1.0, at which no more than 5\% of pairs had two adjacent contacts at or above it. This gave $\tau_\Delta=1.75$, at which 5 of 145 same-day pairs (3.4\%) fired.

\subsection{Evaluation}
\label{sec:eval}

$T$ and $\TD$ were compared on the 68 ears of the evaluation set by agreement, sensitivity and specificity (taking $\TD$ as reference), positive and negative predictive value, Cohen's $\kappa$ and a two-sided Fisher exact test. Threshold-free discrimination was summarized by the area under the receiver operating characteristic curve (AUC) of $-S$ against the $\TD$ label.

To estimate performance on patients not used for calibration, the cohort was split at random into two halves of patients 200 times. In each split the threshold was re-derived on the growth-free ears of one half and applied to the other half. Doing this in both directions gave 400 held-out estimates of agreement with $\TD$ and of the false-positive rate on growth-free ears. $\tau_\Delta$ does not depend on $T$ and was not re-derived. 

Two sensitivity analyses addressed choices in the reference and in calibration. First, $\TD$ was recomputed without the persistence requirement from the intra-operative and last recordings only. Second, because intra-operative recordings are noisier than later ones (Section~\ref{sec:res-robust}), the growth-free recordings were rescored with each ear's reciprocity residual taken from its first follow-up visit, and the threshold re-derived. 

Ears on which $T$ and $\TD$ disagreed were examined individually. To separate disagreements in magnitude from absence of change, a fixed rule was applied to all 68 ears. An ear showed no change at any visit if $\TD$ was at most 1.30 at every follow-up, the 90th percentile of the same statistic over the same-day null.

\begin{figure}[t]
\centering
\includegraphics[width=0.7\textwidth]{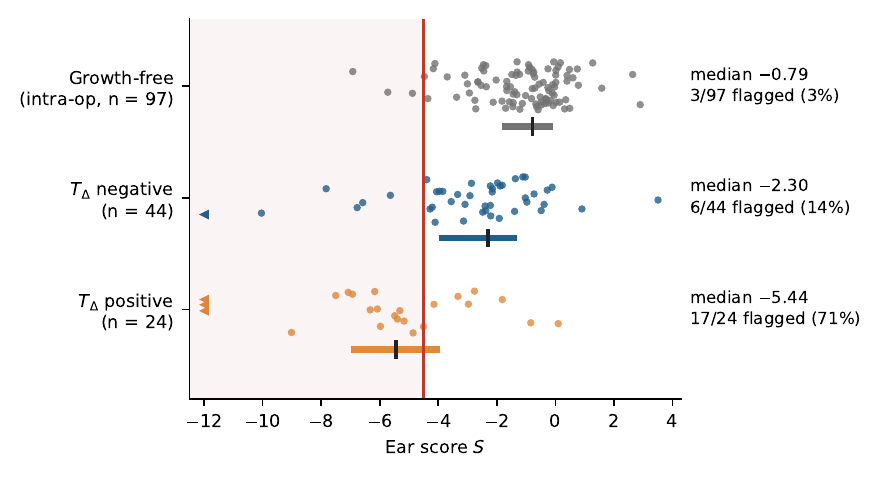}
\caption{Ear scores $S$ for the growth-free intra-operative arm and for the 68 evaluation ears, separated by $\TD$. Bars show the interquartile range and median of each group; the shaded region lies below the threshold $-4.50$, which was set on the growth-free arm alone. Triangles mark scores below $-12$, drawn at the axis limit.}\label{fig:calib}
\end{figure}

\section{Results}
\label{sec:results}

\subsection{Calibration on growth-free recordings}
\label{sec:res-calib}

The calibration rule gave $\tau=-4.50$, at which 3 of the 97 growth-free ears (3.1\%) were flagged (Fig.~\ref{fig:calib}). Growth-free ear scores had a median of $-0.79$ (range $-6.92$ to $2.90$).

\subsection{Agreement with $\TD$}
\label{sec:res-agree}
\begin{figure}[t]
\centering
\includegraphics[width=0.55\textwidth]{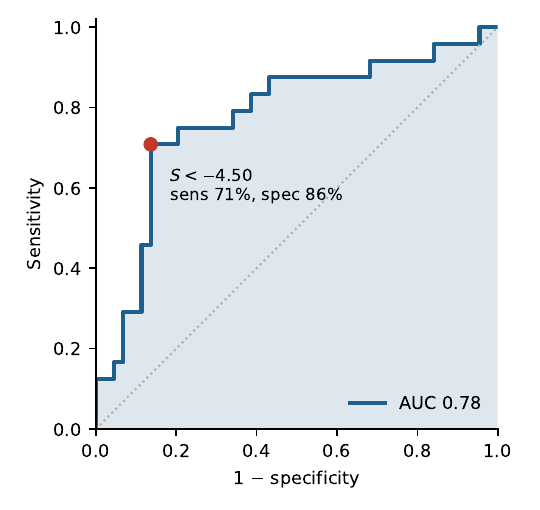}
\caption{Receiver operating characteristic of the ear score $S$ against $\TD$. The shaded area is the AUC, the marker is the calibrated threshold, and the diagonal corresponds to random ranking (AUC 0.5).}\label{fig:auc}
\end{figure}

On the evaluation set, $T$ flagged 23 of 68 ears and $\TD$ 24. The two agreed on 55 ears (17 flagged by both, 38 by neither; 80.9\%), with sensitivity 70.8\% and specificity 86.4\% against $\TD$ ($\kappa=0.58$; $p=5.7\times10^{-6}$; Table~\ref{tab:summary}). Of the ears flagged by $T$, 73.9\% were $\TD$-positive, against a base rate of 35.3\%; of those not flagged, 84.4\% were $\TD$-negative. Across thresholds, $S$ separated $\TD$-positive from $\TD$-negative ears with an AUC of 0.78 (Fig.~\ref{fig:auc}).

With the threshold re-derived on half of the patients over 200 two-fold splits, median held-out agreement was 78.1\% (interquartile range 73.5--81.3\%) at a median false-positive rate of 5.1\% (2.0--8.3\%) on the held-out growth-free ears (Table~\ref{tab:summary}).

\subsection{Sensitivity analyses}
\label{sec:res-robust}

Removing the persistence requirement from $\TD$ changed the reference on 6 of the 68 ears, all of which became positive (30 positive instead of 24), and left agreement with $T$ at 77.9\% (Table~\ref{tab:summary}). 

Intra-operative recordings had a median reciprocity residual of 0.039~mV, against 0.0041~mV for the most recent recordings. Rescoring the growth-free arm with each ear's first follow-up residual gave a threshold of $-4.25$ at the same false-positive count (3/97) and agreement with $\TD$ of 77.9\%. The first follow-up residual exceeded the intra-operative one in 35 of 97 ears, which removed two flags. 

\begin{table}[t]
\caption{Performance of $T$ against $\TD$ under the primary analysis, held-out evaluation, and the sensitivity analyses. PPV: share of ears flagged by $T$ that were $\TD$-positive. FP: false-positive rate on growth-free ears}\label{tab:summary}
\begin{tabular}{@{}lcccccc@{}}
\toprule
Analysis & $n$ & Agree.\ (\%) & Sens.\ (\%) & Spec.\ (\%) & PPV (\%) & FP (\%) \\
\midrule
Primary                               & 68 & 80.9 & 70.8 & 86.4 & 73.9 & 3.1 \\
Two-fold held-out, median             & 34\footnotemark[1] & 78.1 & 71.4 & 82.6 & 69.2 & 5.1 \\
$\TD$ without persistence             & 68 & 77.9 & 63.3 & 89.5 & 82.6 & 3.1 \\
\botrule
\end{tabular}
\footnotetext[1]{Approximately half of the evaluation ears per held-out half; 400 held-out estimates.}
\end{table}

\subsection{Discordant ears}
\label{sec:res-discordant}

$T$ and $\TD$ disagreed on 13 ears (Table~\ref{tab:discordant}), most of them near threshold. Of the 6 ears flagged by $T$ alone, 4 exceeded the 90th percentile of the same-day null at some visit and 3 exceeded $\tau_\Delta$, without persisting over the last two visits. Among the 38 ears negative on both measures, 15 (39\%) exceeded the same 90th percentile and 6 (16\%) exceeded $\tau_\Delta$ without persisting. Two (D4, D6) met the no-change rule, so their flags have no support in the reference. Of the 7 ears positive on $\TD$ alone, 5 scored between $-4.15$ and $-1.81$, below the growth-free median but above the threshold. The other two (D12, D13) were followed for only 193 and 126 days.

\begin{table}[t]
\caption{The 13 ears on which $T$ and $\TD$ disagreed. Max $\TD$: largest single-visit value over all follow-ups. Visits: days after the intra-operative recording of the two visits entering $\TD$. Bold: no change at any visit (Section~\ref{sec:eval})}\label{tab:discordant}
\small
\setlength{\tabcolsep}{4pt}
\begin{tabular}{@{}lrrrc@{\hspace{14pt}}lrrrc@{}}
\toprule
\multicolumn{5}{@{}l}{\emph{Flagged by $T$ only}} & \multicolumn{5}{l@{}}{\emph{Positive on $\TD$ only}} \\
\cmidrule(r{10pt}){1-5}\cmidrule{6-10}
Ear & $S$ & $\TD$ & Max $\TD$ & Visits (d) & Ear & $S$ & $\TD$ & Max $\TD$ & Visits (d) \\
\midrule
D1 & $-12.50$ & 1.19 & 2.14 & 343 / 707 & D7  & $-4.15$ & 2.10 & 2.17 & 511 / 714 \\
D2 & $-10.04$ & 0.70 & 2.09 & 191 / 287 & D8  & $-3.33$ & 2.75 & 4.66 & 344 / 624 \\
D3 & $-7.83$  & 0.87 & 1.56 & 330 / 834 & D9  & $-2.97$ & 2.19 & 2.60 & 140 / 678 \\
\textbf{D4} & {\boldmath$-6.77$} & {\boldmath$-0.13$} & \textbf{0.13} & \textbf{442 / 813} & D10 & $-2.76$ & 2.15 & 3.58 & 581 / 771 \\
D5 & $-6.58$  & 0.67 & 1.93 & 150 / 255 & D11 & $-1.81$ & 2.44 & 2.87 & 365 / 645 \\
\textbf{D6} & {\boldmath$-5.63$} & \textbf{0.25} & \textbf{0.55} & \textbf{672 / 891} & D12 & $-0.84$ & 3.79 & 4.04 & 147 / 193 \\
   &          &      &      &           & D13 & $0.10$  & 3.92 & 4.52 & 57 / 126 \\
\botrule
\end{tabular}
\end{table}

\section{Discussion}

A patient-specific model fitted to apical stimulation predicts the basal SCINSEVs of an ear closely enough that departures from the prediction carry information about growth. Flags raised from a single recording agreed with $\TD$ on 81\% of ears, and on a median of 78\% in held-out patients, while flagging 3--5\% of growth-free ears. A flagged ear was $\TD$-positive in about three cases of four, twice the base rate. Neither the intra-operative baseline nor repeated visits are needed at the time of screening; they enter only the reference against which the screen was evaluated. The fact that EAM misfits correspond to explainable physical changes also indicates that EAM closely captured the underlying physical model and its properties.

The absence of histology is inherent to the problem rather than a gap in this study. In living patients fibrous tissue cannot be seen on clinical CT, and tissue is available only post mortem. $\TD$ is the closest available substitute. It measures persistent change since the intra-operative recording without any model, and its threshold is set on an independent null of same-day repeat recordings. $T$ was calibrated without reference to it. The two measures share the last recording, which $T$ reads against a model prediction and $\TD$ as a change, but they share neither model nor calibration. Agreement between them is evidence that both respond to the same basal process, while disagreements identify where either measure is limited.

Most disagreements were near-threshold: $S$ is a graded score, and several ears missed the threshold by a small margin. A flag indicates electrical signs consistent with basal TG rather than an amount of growth, and is better suited to diagnostic context than to clinical action. Alongside a routine recording, it indicates whether TG is a likely contributor when performance is poorer than expected~\cite{kamakura2016}. Patient-specific uncertainty, for example from the spread of repeated fits, would separate a poor basal fit from a genuine departure and allow a graded measure.

Several limitations should be noted:
\label{sec:discussion-limitations}

\begin{itemize}
\item \textbf{Surrogate reference.} $\TD$ is itself a proxy. Persistence may miss growth that is recent or regressing, and early postoperative change may register as growth (ears D12 and D13).

\item \textbf{Noise and reference-negative ears.} The growth-free arm is necessarily intra-operative, and these recordings are an order of magnitude less reciprocal than later ones. The higher residual compresses $T$, so the threshold set on them may not transfer exactly to follow-up recordings. The difference in recording conditions also bears on specificity: $\TD$-negative ears scored lower than the growth-free arm (median $-2.30$ against $-0.79$) and 6 of 44 (14\%) were flagged. Rescoring the growth-free arm under the follow-up reciprocity residual moved its median only to $-1.03$. A negative reference thus indicates no persistent change above threshold, not absence of growth. These flags therefore reflect either subthreshold growth or a higher false-positive rate under follow-up conditions than calibration shows.

\item \textbf{Tissue type and model.} Fibrous tissue and new bone both raise resistivity and are not distinguished, and the model uses homogeneous tissue classes~\cite{vu-eam}.

\item \textbf{Single center.} All data came from one center and one device family (Advanced Bionics); the apical--basal assumption and the threshold may not transfer to other arrays or insertion depths.
\end{itemize}
\section{Conclusion}

Tissue growth around cochlear implant electrodes is not observable on routine imaging, and judging a single SCINSEV recording has required a patient-specific reference that is often unavailable. Fitting a CT-based patient-specific model to apical stimulation and scoring the basal misfit relative to the apical misfit provides that reference from the recording itself. With a threshold calibrated for at most 5\% false positives on growth-free recordings (3\% observed), the resulting single-visit flag agreed with persistent multi-visit basal change in 81\% of ears, and in a median of 78\% when calibrated and tested on separate patients. Signs of significant growth can therefore be screened for using the postoperative CT and measurements already collected at every visit, as diagnostic context when performance is poorer than expected. Relating the flag to speech outcomes is the next step toward establishing this use.

\backmatter
\section*{Statements and Declarations}

\subsection*{Funding}
This research was supported by the National Institute on Deafness and Other
Communication Disorders grant number R01DC014037. This work is solely the
responsibility of the authors and does not represent the official views of
the NIH.

\subsection*{Competing Interests}
The authors have no relevant financial or non-financial interests to
disclose.

\subsection*{Ethics Approval}
Data collected in this study were authorized by the Vanderbilt IRB, IRB
Approval Number 241479, Vanderbilt University.

\subsection*{Consent to Participate}
This retrospective study used de-identified data; the requirement for informed consent was waived by the Vanderbilt IRB.

\subsection*{Consent to Publish}
The manuscript contains no individually identifiable data.

\subsection*{Data Availability}
The clinical data analyzed in this study are not publicly available, as they
derive from identifiable clinical imaging and implant recordings. Derived
results are available from the corresponding author on reasonable request,
subject to institutional approval.

\subsection*{Code Availability}
The code is not publicly available due to pending intellectual property
considerations. It is available from the corresponding author on reasonable
request, subject to institutional approval.

\subsection*{Author Contributions}
M.V. conceived the method, performed the analysis and wrote the manuscript. J.N. supervised the work. E.B. developed most of the electro-anatomical model code underlying the optimization pipeline and reviewed the manuscript. J.R. set up the tissue class labeling used to build the tissue maps. All authors read and approved the final manuscript.



\begin{thebibliography}{99}

\bibitem{ci-outcomes}
Boisvert I, Reis M, Au A, Cowan R, Dowell RC (2020) Cochlear implantation outcomes in adults: A scoping review. PLOS ONE 15(5): e0232421.
\url{https://doi.org/10.1371/journal.pone.0232421}

\bibitem{somdas2007}
Somdas MA, Li PMMC, Whiten DM, Eddington DK, Nadol JB Jr (2007) Quantitative
evaluation of new bone and fibrous tissue in the cochlea following cochlear
implantation in the human. Audiol Neurotol 12(5):277--284.
\url{https://doi.org/10.1159/000103208}

\bibitem{foggia2019}
Foggia MJ, Quevedo RV, Hansen MR (2019) Intracochlear fibrosis and the foreign
body response to cochlear implant biomaterials. Laryngoscope Investig
Otolaryngol 4(6):678--683.

\bibitem{ishai2017}
Ishai R, Herrmann BS, Nadol JB Jr, Quesnel AM (2017) The pattern and degree of
capsular fibrous sheaths surrounding cochlear electrode arrays. Hear Res
348:44--53.

\bibitem{kamakura2016}
Kamakura T, Nadol JB Jr (2016) Correlation between word recognition score and
intracochlear new bone and fibrous tissue after cochlear implantation in the
human. Hear Res 339:132--141.

\bibitem{wilk2016}
Wilk M, Hessler R, Mugridge K, Jolly C, Fehr M, Lenarz T, Scheper V (2016)
Impedance changes and fibrous tissue growth after cochlear implantation are
correlated and can be reduced using a dexamethasone eluting electrode. PLoS
ONE 11(2):e0147552.

\bibitem{ct-tg}
Heutink F, Klabbers TM, Huinck WJ, Lucev F, van der Woude WJ, Mylanus EAM, Verbist BM (2022)
Ultra-high-resolution CT to detect intracochlear new bone formation after cochlear
implantation. Radiology 302(3):605--612.
\url{https://doi.org/10.1148/radiol.211400}

\bibitem{derijk2020}
de Rijk SR, Tam YC, Carlyon RP, Bance ML (2020) Detection of extracochlear
electrodes in cochlear implants with electric field imaging/transimpedance
measurements: a human cadaver study. Ear Hear 41(5):1196--1207.

\bibitem{eam-prior}
Malherbe TK, Hanekom T, Hanekom JJ (2016) Constructing a three-dimensional
electrical model of a living cochlear implant user's cochlea. Int J Numer
Methods Biomed Eng 32(7):e02751. \url{https://doi.org/10.1002/cnm.2751}

\bibitem{bratu-lineage}
Bratu EL, Berg KA, DeFreese AJ, Gifford RH, Noble JH (2024) An adaptable model
for estimating patient-specific electrical properties of the implanted cochlea.
Proc SPIE 12928, Medical Imaging 2024: Image-Guided Procedures, Robotic
Interventions, and Modeling, 129280Q.
\url{https://doi.org/10.1117/12.3008731}

\bibitem{neural-health}
Liu Z, Cakir A, Noble JH (2020) Auditory nerve fiber health estimation using
patient specific cochlear implant stimulation models. SASHIMI (Workshop) vol. 12417 (2020): 184-194. \url{https://doi.org/10.1007/978-3-030-59520-3_19}

\bibitem{vu-eam}
Vu M, Bratu E, Noble J. (2026) Automated optimization of patient-specific electro-anatomical models fitted to cochlear implant SCINSEV. arXiv:2610.41054, 2026, doi: 10.48550/arXiv.2610.41054. \url{https://arxiv.org/abs/2610.41054}


\bibitem{atlas-seg}
Noble JH, Labadie RF, Majdani O, Dawant BM (2011) Automatic segmentation of
intracochlear anatomy in conventional CT. IEEE Trans Biomed Eng
58(9):2625--2632. \url{https://doi.org/10.1109/TBME.2011.2160262}

\bibitem{elec-localization}
Zhao Y, Chakravorti S, Labadie RF, Dawant BM, Noble JH (2019) Automatic
graph-based method for localization of cochlear implant electrode arrays in
clinical CT with sub-voxel accuracy. Med Image Anal 52:1--12.
\url{https://doi.org/10.1016/j.media.2018.11.005}

\bibitem{bayesopt}
Jones DR, Schonlau M, Welch WJ (1998) Efficient global optimization of
expensive black-box functions. J Glob Optim 13(4):455--492.
\url{https://doi.org/10.1023/A:1008306431147}

\bibitem{cupy}
Okuta R, Unno Y, Nishino D, Hido S, Loomis C (2017) CuPy: a NumPy-compatible
library for NVIDIA GPU calculations. In: Proceedings of the Workshop on
Machine Learning Systems (LearningSys), 31st Conference on Neural Information
Processing Systems (NIPS).
\url{http://learningsys.org/nips17/assets/papers/paper_16.pdf}

\end{thebibliography}
\end{document}